\documentclass[twocolumn,showpacs,amsmath,amssymb]{revtex4}
\usepackage{graphicx}
\begin{document}
\title {\Large \bf Gyromagnetic Ratio of Charged Kerr-Anti-de Sitter Black Holes}
\author{\large Alikram  N. Aliev}
\address{Feza G\"ursey Institute, P.K. 6  \c Cengelk\" oy, 34684 Istanbul, Turkey}
\date{\today}
\begin{abstract}
We examine the gyromagnetic ratios of rotating and charged AdS
black holes in four and higher spacetime dimensions. We compute
the gyromagnetic ratio for Kerr-AdS black holes with an arbitrary
electric charge in four dimensions and show that it corresponds to
$ g=2 $ irrespective of the AdS nature of the spacetime. We also
compute the gyromagnetic ratio for Kerr-AdS black holes with a
single angular momentum and with a test electric charge in all
higher dimensions. The gyromagnetic ratio crucially depends on the
dimensionless ratio of the rotation parameter to the curvature
radius of the AdS background. At the critical limit, when the
boundary Einstein universe is rotating at the speed of light, it
exhibits a striking feature leading to $ g=2 $ regardless of the
spacetime dimension. Next, we extend our consideration to include
the exact metric for five-dimensional rotating charged black holes
in minimal gauged supergravity. We show that  the value of the
gyromagnetic ratio found  in the ``test-charge"  approach remains
unchanged for these black holes.

\end{abstract}


\maketitle
\begin{center}
{\large \bf Introduction}
\end{center}

Black holes continue to play a profound  role in the fundamental
description of nature. The striking examples of this come from the
AdS/CFT correspondence between a weakly coupled gravity system in
an anti-de Sitter (AdS) background and a strongly coupled
conformal field theory (CFT) living on its boundary \cite{mal}.

The simple AdS black hole is described  by the familiar
Schwarzschild-AdS  solution.  The most important feature of the
black hole is that it has  a  minimum critical temperature
determined by the curvature radius of the AdS background. This
means that there must be a thermal phase transition between AdS
space and Schwarzschild-AdS  space at a critical temperature: At
low temperatures thermal radiation in the AdS space is in stable
equilibrium, while at temperatures higher than the critical value
there is no stable equilibrium configuration without a black hole
\cite{hpage}. The Hawking-Page transition was interpreted by
Witten \cite{witten} in terms of a transition between the
confining and deconfining phases of the corresponding conformal
field theory. In further developments, this result has been
extensively discussed in the context of both static and rotating
AdS black holes in various spacetime dimensions (see Refs.
 \cite{emparan2}-\cite{odin}). The authors of work in \cite{hhtr}
have studied the relationship between Kerr-AdS black holes in the
bulk and conformal field theory living in a boundary Einstein
universe. It has been shown that at the critical limit, at which
the boundary Einstein universe is rotating  at the speed of light,
generic thermodynamic features of the conformal field theory agree
with those of the black holes in the bulk.

In general, the clearest description of the boundary conformal
field theory in a rotating Einstein universe imprinted by a
Kerr-AdS black hole in the bulk is a very complicated and subtle
question. Exploring the critical limit at which the Einstein
universe rotates at the speed of light makes a significant
simplification as it incorporates generic features of both bulk
and boundary theories. There are subtleties even with the
definition of the total mass and angular velocities of the
Kerr-AdS black holes. In Ref. \cite{gpp1}, it has been argued that
one must evaluate the mass and angular velocities relative to a
frame which is nonrotating at infinity. Only these quantities
define the most important characteristics of the Kerr-AdS black
holes relevant for their CFT duals and satisfy the first law of
thermodynamics. A more detailed analysis has led to a general
equivalence between the bulk and the boundary thermodynamic
variables \cite{gpp2,gpp3}, thereby clarifying the lack of
unanimity in previous cases  \cite{bala1}.

In the light of these considerations, we address further important
properties of the black holes in AdS space, namely the
gyromagnetic ratios of charged Kerr-AdS black holes. We recall that in
classical electrodynamics the gyromagnetic ratio of a charged
rotating body is always $ g=1 $ for a constant ratio of the charge
to mass density. However, quantum electrodynamics predicts, up to
radiative corrections, that $ g=2 $ for charged fermions, like
electrons and muons. It has been shown that at the tree-level, $
g=2 $ is the natural value of the gyromagnetic ratio for
elementary particles of arbitrary spin \cite{fmt}. The exact
result $ g=2 $ is related to unbroken supersymmetry and the factor
$ g-2 $ is considered to be a measure of SUSY-breaking effects
(see Ref. \cite{dom} for a recent review). It is long known that,
unlike a uniformly charged rotating body, the gyromagnetic ratio
for a rotating asymptotically flat black hole in four-dimensional
Einstein-Maxwell theory is equal to $ 2 $, just like for an
electron \cite{carter1}. The parameter $ g $ is defined as a
constant of proportionality in the equation for the magnetic
dipole moment
\begin{equation}
\mu= g\, \frac{Q\,J}{2 \,M}\,\,, \label{g0}
\end{equation}
where $\,M\,$ is the mass, $\,J\,$ is the angular momentum and
$\,Q\,$ is the electric charge of the black hole.

In recent works \cite{af,aliev}, the gyromagnetic ratio was
studied in higher dimensions for asymptotically flat Myers-Perry
black holes \cite{mp} carrying a test electric charge as well as
for arbitrary values of the electric charge in the limit of slow
rotation. A numerical treatment was given in Refs. \cite{kunz}. It
should be noted that, unlike four dimensions, the value of the
gyromagnetic ratio is not universal in higher dimensions. For a
five-dimensional Myers-Perry black hole with a test electric
charge the gyromagnetic ratio was found to be $ g=3 $. On the
other hand, it is known that for black holes in five-dimensional
Kaluza-Klein theory the $g$-factor approaches unity in the
ultrarelativistic limit \cite{gw}. This value is the natural
$g$-factor for massive states in the Kaluza-Klein theory
\cite{hosoya}. It is also known that for Kaluza-Klein black holes
in ten-dimensional supergravity $ g=1 $ \cite{duff}, while some $
p\,$-brane solutions in higher dimensions have a gyromagnetic
ratio that corresponds to $ g=2 $ \cite{bala2}.

The purpose of this paper is to examine the gyromagnetic ratios
of rotating and charged AdS black holes in four and higher
spacetime dimensions. We begin with the Kerr-Newman-AdS solution
in four dimensions and show that its gyromagnetic ratio
corresponds to $ g=2 $. Since, the counterpart of the
Kerr-Newman-AdS solution in higher dimensions is not known yet, we
employ the test-charge  approach. This allows us to use the
higher-dimensional Kerr-AdS metrics found in Refs. \cite{hhtr,
glpp1} and  propose an elegant way of describing the associated
test electromagnetic field. We focus on the Kerr-AdS
black holes with a single angular momentum which is of crucial
importance in the sense that it allows a universal description of
the gyromagnetic ratio in four and higher dimensions. (Multiple
angular momenta case, by its very existence, requires  higher
dimensions  and  does not admit a passage to the value of the
gyromagnetic ratio in four dimensions).
We find a remarkable formula for the gyromagnetic ratio
\begin{eqnarray}
g &=&  2+ (N-3)\,\Xi\, \,, \label{gyro}
\end{eqnarray}
where $\, \Xi=1 - a^2\,l^{-2} \,$ and $ N $ is the number of
spatial dimensions. Furthermore,  $ a $ is the rotation parameter
and $ l $ is the curvature radius of the AdS background. For $ N=3
$ this expression shows that  $ g=2 $ is a universal feature of
four dimensions irrespective of the AdS nature of the spacetime.
For vanishing cosmological constant, $ l\rightarrow\infty $, it
recovers the value of the gyromagnetic ratio found for weakly
charged Myers-Perry black holes \cite{af,aliev}. However, the most
striking feature of the expression (\ref{gyro}) appears in the
critical limit $ \Xi \rightarrow 0 $, in which the boundary
Einstein universe rotates at the speed of light. We see that $ \,
g\rightarrow 2 \,$ regardless of the spacetime dimension. In a
recent work, Ref. \cite{cvetic1}, it has been shown that at the
critical limit of rotation the Kerr-AdS black holes are related to
SUSY configurations. Thus, it also follows from our result that a
supersymmetric black hole in an AdS background must have the value
of the gyromagnetic ratio  corresponding  to $ g=2 $.

In the final section we go beyond the test-charge approximation and
use the exact metric for five-dimensional rotating charged black holes
in minimal gauged supergravity that was recently found in \cite{cvetic2}.
Focusing on the case with a single rotation parameter, we show that the
gyromagnetic ratio of these black holes is the same as that appearing
in (\ref{gyro}) for $ N=4 $.

\begin{center}
{\bf \large Kerr-Newman-AdS black holes}
\end{center}
The exact solution describing the Kerr-Newman-AdS black holes in
four-dimensional Einstein-Maxwell theory was found in
\cite{carter2}. It is given by the metric
\begin{eqnarray}
ds^2 & = &-{{\Delta_r}\over {\Sigma}} \left(dt - \frac{a
\sin^2\theta}{\Xi}\,d\phi\,\right)^2 + {\Sigma \over~ \Delta_r}
dr^2 + {\Sigma \over ~\Delta_{\theta}}\,d\theta^{\,2} \nonumber \\
& &  + \frac{\Delta_{\theta}\sin^2\theta}{\Sigma} \left(a\, dt -
\frac{r^2+a^2}{\Xi} \,d\phi \right)^2 \,\,, \label{4kads}
\end{eqnarray}
where $ {\Xi} $ has the same form as that appearing in
(\ref{gyro}) and
\begin{eqnarray}
\Delta_r &= &\left(r^2 + a^2\right)\left(1 +\frac{r^2}{l^2}\right)
- 2 M r +Q^2 \,, \\
\Delta_\theta & =&  1 -\frac{a^2}{l^2} \,\cos^2\theta\,,~~ \Sigma
= r^2+ a^2 \cos^2\theta \,.
\end{eqnarray}
The associated electromagnetic field is described by the potential
one-form
\begin{equation}
A= -\frac{Q r}{\Sigma} \left(dt- \frac{a \sin^2\theta}{\Xi}\,d\phi
\right)\,\,, \label{potform1}
\end{equation}
where  the parameter $ Q  $ is related to the electric charge by Gauss's law which gives
\begin{equation}
Q^{\,\prime}=  \frac{Q}{\Xi} \,\,.\label{charge}
\end{equation}
Here and in the following the primed quantities refer to the
physical characteristics of the black holes. The parameters $ M $
and  $ a $ appearing in the metric (\ref{4kads}) are related to
its total mass and angular momentum which are given by the first
law of thermodynamics \cite{gpp1}
\begin{eqnarray}
M^{\prime}&=& \frac{M}{\Xi^2}\,\,,~~~~~~ J^{\prime}= \frac{a
M}{\Xi^2}\,\,.\label{tmj1}
\end{eqnarray}
Clearly, a rotating charged AdS black hole must have a magnetic
dipole moment. The most direct way to determine it is to examine
the asymptotic behavior of the magnetic field generated by the
black hole. For this purpose, it is useful to introduce an
orthonormal frame which is given by the basis one-forms
\begin{eqnarray}
\label{basis}
 e^{0} &=&\left(\frac{\Delta_r }{\Sigma
}\right)^{1/2}\left(dt- \frac{a \sin^2\theta}{\Xi }\,d\phi
\right)\,\,,\nonumber \\[2mm]
e^{3} &=&\left(\frac{\Delta_{\theta} }{\Sigma }\right)^{1/2}\,\sin
\theta \left(a \,dt- \frac{r^2+a^2}{\Xi }\,d\phi \right)\,, \nonumber \\[2mm]
e^{1}& =& \left(\frac{\Sigma}{\Delta_r }\right)^{1/2}dr\,\,,
~~~~~e^{2}= \left(\frac{\Sigma}{\Delta_{\theta }}\right)^{1/2}
d\theta  \,.
\end{eqnarray}
The remarkable property of this frame is that an observer at rest
in it measures only the radial components of the electric and
magnetic fields. This is confirmed by calculating the
electromagnetic field two-form  in this frame
\begin{equation}
 F=-\frac{Q}{\Sigma^2}\,
 \left[ \left(2 r^2 - \Sigma \right) e^{0}\wedge e^{1}
 -
 2 a\,  r \cos\theta \,e^{3}\wedge e^{2}
 \right]\,.
\label{2form}
\end{equation}
From the asymptotic behavior of this expression it is easy to read
off both the electric charge given in (\ref{charge}) and the
magnetic dipole moment
\begin{eqnarray}
\mu ^{\,\prime}&=&Q^{\,\prime}a=\frac{\mu}{\Xi}\,\,,
\label{physmagm}
\end{eqnarray}
where $ \mu= Q a $ is the magnetic dipole moment parameter. The
magnetic dipole moment is related to the mass and angular momentum
in (\ref{tmj1}) by the gyromagnetic ratio. We have a definition
similar to (\ref{g0}) for asymptotically flat black holes; that is
\begin{equation}
\mu^{\,\prime}= g\, \frac{ Q^{\,\prime}J^{\,\prime}}{2
M^{\,\prime}}\,\,. \label{g}
\end{equation}
It follows that the Kerr-Newman-AdS  black holes must have a
gyromagnetic ratio $ g=2 $ just as the usual Kerr-Newman black
holes in asymptotically flat spacetime.
\begin{center}
{\large \bf Higher dimensional  case}
\end{center}
The Kerr-AdS black holes with a single rotation parameter in $ N+1
$ dimensions are described by the metric
\begin{eqnarray}
ds^2 & = & -{{\Delta_r}\over {\Sigma}} \left(dt - \frac{a
\sin^2\theta}{\Xi}\,d\phi\,\right)^2 + {\Sigma \over~ \Delta_r}
dr^2 \nonumber \\ && + {\Sigma \over
~\Delta_{\theta}}\,d\theta^{\,2} \nonumber +
\frac{\Delta_{\theta}\sin^2\theta}{\Sigma} \left(a dt -
\frac{r^2+a^2}{\Xi} \,d\phi \right)^2
\\[2mm] &&
+ \,r^2 \cos^2{\theta} \, d\Omega_{N-3}^2\,,\label{hkads}
\end{eqnarray}
where the metric functions are the same as given in (\ref{4kads})
except for
\begin{eqnarray}
\Delta_r &= &\left(r^2 + a^2\right)\left(1 +\frac{r^2}{l^2}\right)
- m \,r^{4-N} \,
\end{eqnarray}
and  $~ d\Omega_{N-3}^2\, $ is the metric on a unit $(N-3)
$-sphere. Here $ m $ is the mass parameter that reduces to $ 2 M $
for $ N=3 $.

Let us now assume that a generic Kerr-AdS black hole in higher
dimensions \cite{glpp1} may carry a test electric charge. It is
straightforward to show that {\it the Killing one-form $ \delta
\hat\xi_{(t)} $ that represents the difference between the
timelike generators in the spacetime metric  and in its reference
background can be used as a potential one-form for the test
electromagnetic field of the black hole.} Using this prescription for
the metric (\ref{hkads}), we find the potential one-form
\begin{equation}
A= -\frac{Q\,r^{4-N}}{(N-2)\,\Sigma}\left(dt- \frac{a
\sin^2\theta}{\Xi}\,d\phi \right)\,\,, \label{hpotform}
\end{equation}
satisfying the Maxwell equations. The Gauss law gives rise to the
same relation for the electric charge as that given in
(\ref{charge}). For $ N=3 $ equation (\ref{hpotform}) agrees  with
(\ref{potform1}).

The expressions for the total mass and angular momenta of general
Kerr-AdS metrics in higher dimensions which are consistent with
the first law of thermodynamics were found in \cite{gpp1}.
Specializing these expressions to the case of a single angular
momentum  we obtain
\begin{equation}
M^{\prime}= \frac{m^{\prime} A_{N-1}}{16\pi\,}\,\left[\,2+
(N-3)\,\Xi\, \right],~~
 J^{\prime} = \frac{j^{\,\prime}
A_{N-1}}{8\pi}\,, \label{newmjall}
\end{equation}
where $\,A_{N-1}\,$ is the area of a  unit $\,(N-1)$-sphere and we
have defined the specific mass and angular momentum
\begin{eqnarray}
m^{\prime}&=& \frac{m}{\Xi^2}\,\,,~~~~~~~~ j^{\,\prime}= \frac{a
m}{\Xi^2}\,\,\label{specificmj}
\end{eqnarray}
which are reminiscent of the relations in (\ref{tmj1}).

As in four-dimensional case, the magnetic dipole moment can be
determined from the asymptotic behavior of the electromagnetic
field two-form written in a natural ortonormal frame of the metric
(\ref{hkads}), in which it takes the simplest form
\begin{eqnarray}
\label{2formall}
 F&=&- \frac{Q\,r^{3-N}}{(N-2)\,\Sigma{\,^2}} \,\left\{\, \left[(N-2) \,\Sigma -2\, a^2
 \cos^2\theta\right]
 \right.  \nonumber \\[2mm]  & & \left.
 \,e^{0}\wedge e^{1}
 -2 a\, r \cos\theta \,e^{3}\wedge e^{2}
 \right\}\,\,.
\end{eqnarray}
It follows that the dominant behavior of the magnetic field at
spatial infinity is determined by the magnetic dipole moment
\begin{eqnarray}
\mu^{\,\prime}&=&\frac{j^{\,\prime}\,Q^{\,\prime}}{m^{\,\prime}}\,
= \left[\,2+ (N-3)\,\Xi\, \right]
\frac{J^{\,\prime}\,Q^{\prime}}{2\,M^{\prime}}\,\,. \label{gyro1}
\end{eqnarray}
Comparing now this expression with (\ref{g}) we read off the value
of the gyromagnetic ratio given in equation (\ref{gyro}) .

It is important to note that the value of the gyromagnetic ratio
found above can be proved by alternative calculations employing a
distinct approach. We define the twist of the Killing one-form
$\,\hat \xi_{(t)} $ that is given by the  $(N-2)$-form
\begin{equation}
{\hat \omega}_{N-2}  =
\frac{1}{(N-2)}\,^{\star}\left({\hat\xi}_{(t)} \wedge
d\,{\hat\xi}_{(t)}\right)\,. \label{twist1}
\end{equation}
Physically, this quantity measures the failure of the Killing
vector $ \partial_{t} $  to be hypersurface orthogonal. The $
{\star} $ operator denotes the Hodge dual. Evaluating the twist
form (\ref{twist1}) in the metric (\ref{hkads}) we find that it is
closed, i.e. $d {\hat \omega}_{N-2}=0 $. This implies  the
existence (locally) of the twist potential $ (N-3) $-form which
after performing the background subtraction is given by
\begin{equation}
\delta \hat\Omega_{N-3}  = \frac{a m}{\Sigma}
\,\,\frac{\cos^{N-2}\theta}{N-2}\,\, d\Sigma_{N-3}\,\,,
\label{twistpot3}
\end{equation}
where $ d\Sigma_{N-3} $ is defined on a unit $(N-3)$-sphere. Next,
we  define the magnetic field $ (N-2)$-form
\begin{eqnarray}
{\hat B}_{N-2} &=& i_{\hat\xi_{(t)}}\, ^{\star}F =\,
^{\star}\left({\hat\xi_{(t)}}\wedge F\right)\,\,, \label{mform}
\end{eqnarray}
which in the metric (\ref{hkads}) can be expressed as
\begin{equation}
{\hat B}_{N-2}= -d\,\varphi_{N-3} \,\,, \label{magpot}
\end{equation}
where the magnetic potential $ (N-3) $-form
\begin{equation}
\varphi_{N-3} = \frac{a Q}{\Sigma}
\,\,\frac{\cos^{N-2}\theta}{N-2}\,\, d\Sigma_{N-3}\,\,.
\label{varphi}
\end{equation}
We see that the asymptotic behavior of this expression determines
the magnetic dipole moment parameter $ \mu = Q a $ just as the
twist potential $ (N-3)$-form  in (\ref{twistpot3}) determines the
specific angular momentum parameter $ j=a m  $. This fact can also
be expressed in the form
\begin{equation}
\varphi_{N-3} =  \frac{Q}{m}\,\delta \hat\Omega_{N-3} \,\,,
\label{proofrel}
\end{equation}
which is equivalent to the relation (\ref{gyro1}). This proves the
value of the gyromagnetic ratio in (\ref{gyro}).

\begin{center}
{\large \bf Five-dimensional black holes in minimal gauged supergravity }
\end{center}

In obtaining the gyromagnetic ratio in (\ref{gyro}) we have used
the test-charge approximation and thermodynamically consistent
expressions for the mass and angular momentum. Now, it is natural
to ask how this result will change for an arbitrary electric
charge of the black hole.  First of all, one may expect that
in general the gyromagnetic ratio will depend on the value of
the electric charge. However, in some cases the formula (\ref{gyro})
may still work (at least in the critical limit of rotation
$  \Xi \rightarrow 0 $) for the arbitrary electric charge.
As an illustration of this idea, we calculate the gyromagnetic ratio for  rotating  charged black holes in minimal five-dimensional  gauged supergravity with the Lagrangian
\begin{eqnarray}
{\cal L} &=& (R+12/l^2)\,^{\star} 1 -\frac{1}{2} \,\, ^{\star}F\,\wedge F \nonumber \\ &&
+\frac{1}{3\sqrt{3}}\,\, F\,\wedge F\,\wedge A\,\,,
\label{5dlag}
\end{eqnarray}
where  $ F=d\,A $. The exact metric for these black holes with two independent rotation parameters was found in \cite{cvetic2}. Specializing this metric to the case of a single rotation parameter and using the Boyer-Lindquist type coordinate system that is rotating at spatial infinity we obtain that
\begin{eqnarray}
\label{5dkads}
ds^2 & = & - f \left(dt - \frac{a
\sin^2\theta}{\Xi}\,d\phi +\frac{Q a \cos^2\theta}{f\,\Sigma} \,d\psi \right)^2  \nonumber \\ &&
+\, \frac{\Delta_{\theta}\sin^2\theta}{\Sigma} \left(a dt -
\frac{r^2+a^2}{\Xi} \,d\phi \right)^2  \nonumber \\ &&
+\, \cos^2\theta \left(r^2+ \frac{Q^2 a^2 \cos^2\theta}{f \Sigma^2} \right) d\psi^2 \nonumber \\ &&
+ \,{\Sigma \over~ \Delta_r} dr^2  + {\Sigma \over
~\Delta_{\theta}}\,d\theta^{\,2}\,\,,
\label{5dkads}
\end{eqnarray}
where
\begin{eqnarray}
\Delta_r &= &\left(r^2 + a^2\right)\left(1 +\frac{r^2}{l^2}\right) +\frac{Q^2}{r^2}- m \,\,,\\
f& =&  {{\Delta_r}\over {\Sigma}} + \frac{Q^2}{\Sigma^2}- \frac{Q^2}{r^2 \Sigma}\,\,.
\end{eqnarray}
We note that under the coordinate transformation $\phi \rightarrow \phi- (a/l^2)\, dt $, the metric (\ref{5dkads}) reduces to the form given in \cite{cvetic2}.
The corresponding  potential one-form is given by
\begin{equation}
A= -\frac{\sqrt{3}\, Q }{\Sigma} \left(dt- \frac{a \sin^2\theta}{\Xi}\,d\phi
\right)\,\,, \label{5dpotform2}
\end{equation}
where the parameter  $ Q  $ is related to the electric charge by the generalized Gauss law
\begin{equation}
Q^{\,\prime}= {\frac{1}{2 \pi^2}} \oint \,\left(^{\star}F -F\wedge A/\sqrt{3}\right)
\label{gauss}
\end{equation}
giving
\begin{equation}
Q^{\,\prime}=  \frac{2 \sqrt{3}\,Q}{\Xi} \,\,.\label{5dcharge}
\end{equation}

It is important to note that  though the metric  (\ref{5dkads}) has a single rotation parameter, evaluating  the Komar integrals
\begin{eqnarray}
J^{\prime}_{\phi}& = & \frac{1}{16 \pi\,}\oint \,^{\star}d\hat
\xi_{(\phi)}\,,~~~J^{\prime}_{\psi} =  \frac{1}{16 \pi\,}\oint \,^{\star}d\hat \xi_{(\psi)}\,\,,\label{komar}
\end{eqnarray}
where the Killing one-form $\,\hat \xi=\xi_{\mu}\, d x^{\mu}$ is associated
with the Killing vector  $\xi_{(\phi)}= \partial/\partial \phi\,,$ or
$\xi_{(\psi)}= \partial/\partial \psi\,$, we find two angular momenta
\begin{eqnarray}
J^{\prime}_{\phi} &= &\frac{\pi}{4}\,\frac{a m}{ \Xi^2}\,,
~~~~J^{\prime}_{\psi} = \frac{\pi}{4}\,\frac{Q a }{\Xi}\,\,.
 \label{5dj}
 \end{eqnarray}
That is, in the absence of the second rotation parameter in five dimensions, the metric (\ref{5dkads}) still possesses an additional (induced) angular momentum in $ \psi $-direction due to its electric charge (see also \cite{cvetic2}). The total mass  of the metric that is consistent with the first law of thermodynamics is given by the expression
\begin{equation}
M^{\prime}= \frac{\pi m}{8\,\Xi^2}\,\left(\,2+
\Xi\, \right)\,\,,
\label{5dm}
\end{equation}
which agrees with (\ref{newmjall}) for $ N=4 $.

Next, we determine the magnetic dipole moment of the black hole. As in the previous  cases above, we first examine the asymptotic behavior of the electromagnetic field in  an orthonormal frame with the basis one-forms
\begin{eqnarray}
\label{basisall}
 e^{0} &=& f ^{1/2}\,\left(dt - \frac{a
\sin^2\theta}{\Xi}\,d\phi +\frac{Q a \cos^2\theta}{f\,\Sigma} \,d\psi \right) \,, \nonumber \\[2mm]
e^{3} &=&\left(\frac{\Delta_{\theta} }{\Sigma }\right)^{1/2}\,\sin
\theta \left(a \,dt- \frac{r^2+a^2}{\Xi }\,d\phi
\right)\,\,, \nonumber\\[2mm]
e^{4} & =& r \cos\theta\,\left(1+ \frac{Q^2 a^2 \cos^2\theta}{f r^2 \Sigma^2}
\right)^{1/2} d\psi \,\,, \nonumber\\[2mm]
e^{1} &=&\left(\frac{\Sigma}{\Delta_r
}\right)^{1/2}dr\,\,,~~~~~e^{2}=
\left(\frac{\Sigma}{\Delta_{\theta }}\right)^{1/2} d\theta
\,\,.
\end{eqnarray}
This is a generalization of the natural orthonormal frame (\ref{basis}) to include the metric (\ref{5dkads}).
The electromagnetic field two-form written in this  frame takes its simplest form
\begin{eqnarray}
\label{5d2form}
 F&=&- \frac{2 \sqrt{3}\,Q}{\Sigma{\,^2}}\,\left( {{\Delta_r}\over {f \Sigma}}\right)^{1/2}
 \,\left[ r \,e^{0}\wedge e^{1}
\right.  \nonumber \\[2mm]  & & \left.
+ \,\frac{Q a \cos\theta}{f^{1/2} \Sigma}\,\left(1+ \frac{Q^2 a^2 \cos^2\theta}{f r^2 \Sigma^2} \right)^{-1/2} \,e^{1}\wedge e^{4}\right] \nonumber \\ [2mm]  &&
+ \frac{2 \sqrt{3}\,Q a \cos\theta }{\Sigma{\,^2}} \,\,e^{3}\wedge e^{2}
 \,\,.
\end{eqnarray}
We note that this expression, unlike (\ref{2formall}), involves  a higher-order in $ 1/r $ term (the second term in square parentheses) that  appears due to the induced angular momentum in the $ \psi $-direction. For the dominant behavior of these fields at spatial infinity we find that
\begin{eqnarray}
\label{5d asymptotic1}
F_{\hat{r}\hat{t}}& = &\frac{Q^{\,\prime}\,\Xi}
{r^3}
+\mathcal{O}\left(\frac{1}{r^{5}}\right) \,\,,\\[2mm]
\label{5d asymptotic2}
F_{\hat{\theta}\hat{\phi}} &= &\frac{Q^{\,\prime} a\,\Xi \cos\theta}{r^4}
+\mathcal{O}\left(\frac{1}{r^{6}}\right) \,\,,\\[2mm]
F_{\hat{r}\hat{\psi}}& = & \mathcal{O}\left(\frac{1}{r^{6}}\right)\,\,.
\label{5d asymptotic3}
\end{eqnarray}

It is easy to check that the Gauss flux of the radial electric field (\ref{5d asymptotic1}) confirms the value of the electric charge in  (\ref{5dcharge}). From  the asymptotic behavior of the magnetic field in Eq.(\ref{5d asymptotic2}), it follows that the magnetic dipole moment is given by the same relation as in  (\ref{physmagm}); that is,
 \begin{eqnarray}
\mu ^{\,\prime}&=&Q^{\,\prime}a=\frac{\mu}{\Xi}\,\,,
\label{5dphysmagm}
\end{eqnarray}
where we have introduced the magnetic dipole moment parameter $ \mu= 2 \sqrt{3}\, Q a $. The magnetic dipole moment can be written in terms of the total mass, angular momentum (gravitomagnetic dipole moment) and the electric charge of the black hole as follows
\begin{eqnarray}
\mu^{\,\prime}&=&
 \left(\,2+ \Xi\, \right)
\frac{J^{\,\prime}\,Q^{\prime}}{2\,M^{\prime}}\,\,. \label{5dgyrorel}
\end{eqnarray}
From a comparison of this expression with (\ref{g}) we find the gyromagnetic ratio
\begin{equation}
g =  2 + \Xi \,\,.
\label{5dgyro}
\end{equation}
We see that this expression  is exactly the same as that following from (\ref{gyro}) for $ N=4 $. Thus, a rotating charged black hole with a  single rotation parameter in minimal five-dimensional  gauged supergravity  can be assigned the same gyromagnetic ratio as  a five-dimensional Kerr-AdS black hole carrying a test electric charge. That is, in the critical limit of rotation ($  \Xi \rightarrow 0 $) the gyromagnetic ratio $ g \rightarrow 2 $ regardless of black hole's electric charge as well.

\begin{center}
{\large \bf Conclusion}
\end{center}
We have computed the gyromagnetic ratios for rotating and charged
AdS black holes in four and higher dimensions. We have obtained
the exact result $ g=2 $ for the gyromagnetic ratio in four
dimensions, thereby extending the validity of the gyromagnetic
ratio for the usual Kerr-Newman black holes to include  the
Kerr-Newman-AdS black holes. Assuming that generic Kerr-AdS black
holes  in all higher dimensions may carry a test electric charge
we have proposed  an elegant way of describing the electromagnetic
field of these black holes. We have computed the gyromagnetic
ratio of the Kerr-AdS black holes with a single angular momentum
in all higher dimensions. The gyromagnetic ratio  crucially
depends on the dimensionless ratio of the rotation parameter to
the curvature radius of the AdS background. The striking feature
of this dependence appears for maximally rotating black holes $(\Xi \rightarrow 0) $
for which the gyromagnetic ratio approaches $ g=2 $ regardless of the spacetime dimension. In this case the boundary of the AdS spacetime is rotating at the speed of light. These results are of importance for several reasons: Although we do not yet have the exact metrics for charged Kerr-AdS black holes in higher dimensions, we can still examine the electromagnetic properties of these black holes and learn about their gyromagnetic ratio using the test-charge approach.  In addition, they show that $ g=2 $ is
the universal feature of black holes in four dimensions
irrespective of the AdS behaviour the spacetime. As is known, the
exact result $ g=2 $   for elementary particles of arbitrary spin
$ g=2 $ is related to supersymmetry and a deviation from this
value is supposed to be a measure of SUSY-breaking effects
\cite{fmt, dom}. It follows from our result that the value $ g=2 $
for maximally rotating case is also related to supersymmetry,
since in this limit of rotation we have supersymmetric AdS black
holes \cite{cvetic1}. We have also gone beyond the test-charge approximation and computed the gyromagnetic ratio of rotating charged black holes with a  single rotation parameter in minimal five-dimensional gauged supergravity. The gyromagnetic ratio turned out to be the same as for  Kerr-AdS black holes with a test electric charge in five dimensions.

As we have mentioned above, there is a satisfactory agreement between generic thermodynamic features of a Kerr-AdS black hole in the bulk and its CFT dual living on the boundary Einstein space that is rotating at the speed of light \cite{hhtr}. Here, in the critical limit of rotation, we have found that the gyromagnetic ratio $ g=2 $ regardless of spacetime dimension and black hole's electric charge. On this basis and also taking into account the fact that throughout the paper  we have used thermodynamically consistent expressions for the mass and angular momentum, one may conclude that  our result may have a  relevance for AdS/CFT correspondence. That is, one may expect the same value $ g=2 $ for charged conformal matter rotating  at  the speed of light in a boundary Einstein universe.

The author thanks the Scientific and Technological Research
Council of Turkey (T{\"U}B\.{I}TAK) for partial support under the
Research Project 105T437. He also thanks\\ G. W. Gibbons and B.
Mashhoon for useful comments.

\end{document}